\documentclass[fleqn,usenatbib]{mnras}

\usepackage{newtxtext,newtxmath}

\usepackage[T1]{fontenc}
\usepackage{ae,aecompl}

\usepackage{graphicx}	
\usepackage{amsmath}	
\usepackage{amssymb}	

\newcommand{\ME}{M$_{\oplus}$} 
 
\newcommand{\trat}{\chi}

\title[Primordial Entropy of Jupiter]{The Primordial Entropy of Jupiter}

\author[A. Cumming, R. Helled, and J. Venturini]{
Andrew Cumming,$^{1,2}$\thanks{E-mail: andrew.cumming@mcgill.ca}
Ravit Helled,$^{3}$
and Julia Venturini$^{3}$
\\
$^{1}$Department of Physics and McGill Space Institute, McGill University, 3550 rue University, Montreal, QC, H3A 2T8, Canada\\
$^{2}$Institut de Recherche sur les Exoplan\`etes (iREx), Universit\'e de Montr\'eal, C.P. 6128 Succ. Centre-ville, Montreal, QC H3C 3J7, Canada\\
$^{3}$Center for Theoretical Astrophysics and Cosmology, Institute for Computational Science, University of Zurich, Winterthurerstrasse 190, 8057,\\Zurich, Switzerland
}

\date{Accepted XXX. Received YYY; in original form ZZZ}

\pubyear{2018}

\begin{document}
\label{firstpage}
\pagerange{\pageref{firstpage}--\pageref{lastpage}}
\maketitle

\begin{abstract}
The formation history of giant planets determines their primordial structure and consequent evolution. We simulate various formation paths of Jupiter to determine its primordial entropy, and find that a common outcome is for proto-Jupiter to have non-convective regions in its interior. We use planet formation models to calculate how the entropy and post-formation luminosity depend on model properties such as the solid accretion rate and opacity, and show that the gas accretion rate and its time evolution  play a key role in determining the entropy profile. The predicted luminosity of Jupiter shortly after formation varies by a factor of $2$--$3$ for different choices of model parameters. We find that entropy gradients inside Jupiter persist for $\sim 10\ {\rm Myr}$ after formation. 
We suggest that these gradients should be considered together with heavy-element composition gradients when modeling Jupiter's evolution and internal structure.
\end{abstract}




\section{Introduction} \label{sec:intro}

Constraining Jupiter's interior and understanding giant planet formation are major goals in astrophysics and planetary science. For this, we must first understand the primordial structure of giant planets and whether they are fully adiabatic (fully convective). While most interior models of present-day Jupiter are adiabatic \citep[e.g.,][]{miguel16}, recent models suggest that composition gradients and non-convective/layered-convection regions may exist \citep[e.g.,][]{stevenson85,lecontechab12,vazan16,nettel15}. Understanding how much of the interior is convective is crucial in connecting planet formation models to observations of Jupiter today, in particular from Juno (e.g.~\citealt{wahl17}). As well as efficiently transporting heat, convection can also redistribute heavy elements in the planetary interior; in turn, heavy element gradients can shut down convection \citep{vazan16}. Therefore, the planetary cooling rate and the evolution of its internal structure both depend on the internal heat transport mechanism. 

Previous work has shown that Jupiter could be non-adiabatic due to composition gradients laid down in its deep interior during its formation \citep{lozovs17,helledsteven17}. The accretion shock in the final stage of formation also plays a crucial role in setting the entropy profile. In the core accretion model \citep{P96} rapid gas accretion (often referred as phase 3, or runaway gas accretion) occurs once the envelope is relatively massive, and contracts rapidly. Eventually, the gas accretion rate exceeds the rate at which matter can be supplied by the disk, and the planet enters the detached phase in which the gas accretion is hydrodynamic, nearly in free fall, onto a shock at the surface of the protoplanet \citep{Bodenheimer2000}. 
Giant planet formation models typically simplify the treatment of the accretion shock. However, most of the mass is accreted during this stage and the shock's efficiency must be studied properly in order to determine the primordial entropy of giant planets \citep{Marley2007,Chabrier2007,Marleau2017,Mordasini2017}. For massive gas giants, \cite{Berardo2017a} and \cite{Berardo2017b} (hereafter BC17) recently showed that depending on the assumptions made about the accretion shock, the accreted material can have larger entropy than the planet's interior, leading to entropy increasing radially outwards, giving a radiative interior at formation even for a homogeneous composition.

\begin{figure*}
\includegraphics[width=1.8\columnwidth]{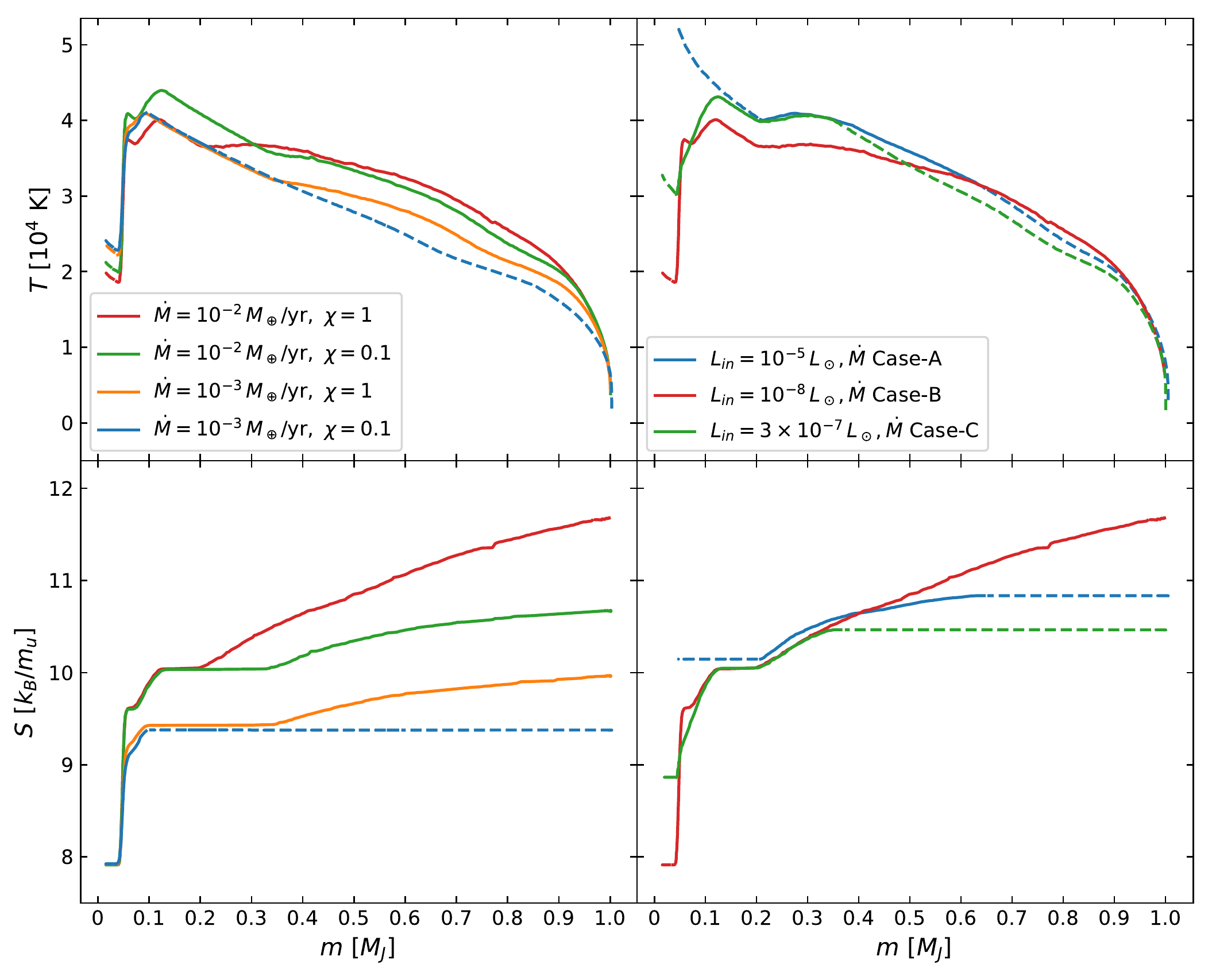}
\caption{Profiles of proto-Jupiter's temperature (top) and entropy (bottom) at the time when accretion ends for different model parameters, with dashed lines indicating convection. {\bf Left:} two different peak accretion rates $\dot M_{\rm max} = 10^{-3}$ and $10^{-2} M_{\oplus}\ {\rm yr^{-1}}$ and shock temperatures corresponding to $\trat=0.1$ and $1$ (see eq.~[\ref{eq:T0}]). The accretion rate determines the overall entropy value, including the plateau in the inner regions, whereas $\trat$ mostly influences the entropy profile in the outer layers. All models have an entropy $\approx 8.0\ k_b/m_u$ at the beginning of the simulation ($L_{\rm in} = 10^{-8} L_\odot$) and accretion Case-B. {\bf Right:} different core luminosities $L_{\rm in}$ and time-dependence of accretion rate, for models with $\dot M_{\rm max} =10^{-2}\, M_{\oplus}\,\mathrm{yr^{-1}}$ and $\trat = 1$. Larger $L_{\rm in}$ leads to a higher starting entropy, driving a larger convective region outside the core. A decreasing accretion rate at later times gives an outer convection zone.}
\label{fig:prof_1}
\end{figure*}

In this paper, we investigate various paths for Jupiter's formation and determine  its primordial entropy and heat transport mechanism. We first extend the study of BC17 to allow a realistic description of Jupiter's formation --- we start the simulation at a much earlier stage (smaller mass) and include a more realistic gas accretion history --- and explore under what conditions Jupiter could have a non-adiabatic interior (\S 2). We then use formation models to determine the range of conditions expected for Jupiter's formation and assess the possibility of proto-Jupiter not being fully convective  (\S 3). In \S 4, we summarize the main results and discuss the evolution post-accretion, including whether radiative regions present at formation persist in Jupiter today.

\section{A radiative or Convective proto-Jupiter?}\label{method1}

Building on the work of BC17, we simulate the growth of Jupiter during the detached phase when H-He gas falls onto an accretion shock at the planet's surface. In this section, we explore the larger parameter space of possibilities without assuming a specific formation model. We start when the H-He envelope has twice the mass of the heavy-element core, and explore the influence of accretion during the detached phase on the final structure. We use the Modules for Experiments in Stellar Astrophysics (MESA) code (version 10108; \citealt{Paxton2011,Paxton2013,Paxton2015,Paxton2017}) to follow the gas envelope.
We assume a homogeneous composition of H-He, with He mass fraction $Y=0.243$, and use the SCVH equation of state \citep{Saumon1995} and dust-free opacities from \cite{Freedman2008}. The inner boundary is at the edge of the core (core density $5\ {\rm g\ cm^{-3}}$).

The entropy of the H-He envelope at the beginning of the detached phase plays an important role in determining whether the inner regions are convective as the planet grows. We set the envelope entropy in the initial model by adjusting the core's luminosity $L_{\rm in}$, and allowing the envelope to reach thermal equilibrium. 
The expected envelope luminosities can be estimated from the accretion luminosity from solids if they reach the core, $L_Z = 8.8\times 10^{-8}\,L_\odot\ (\dot M_Z/10^{-6}\ M_\oplus\ {\rm yr^{-1}})(M_c/5\,M_\oplus)^{2/3}$ for a core density $5\,{\rm g\ cm^{-3}}$. We consider three cases: a $5\,M_\oplus$ core and $10\,M_\oplus$ envelope with (specific) entropies $S = 7.9\,k_b/m_u$ and $8.8\,k_b/m_u$ ($L_{\rm in}=10^{-8}\,L_\odot$ and $3\times 10^{-7}\,L_\odot$), and a $15\,M_\oplus$ core with a $30\,M_\oplus$ envelope and entropy $S = 10.1\,k_b/m_u$ ($L_{\rm in} = 10^{-5}\,L_\odot$). 

We then model gas accretion until the planet reaches Jupiter's mass. The temperature $T_0$ of the accreted material after passing through the accretion shock is uncertain. Recent radiation-hydrodynamic simulations by \cite{Marleau2017} indicate that the shock is a supercritical shock in which almost all of the accretion luminosity $GM\dot M/R$ is radiated away and the shock is isothermal. However, depending on how much of the luminosity is absorbed by the accretion flow and reaccreted, some fraction may be incorporated into the planet. The shock temperature $T_0$ scales with accretion rate, planet mass and radius as $T_0\propto T_{\rm accr}$ (\citealt{Marleau2017}, see also \citealt{Stahler1980}), where $T_{\rm accr}$ is defined by $4\pi R^2 \sigma T_{\rm accr}^4 = GM\dot M/R$, giving
\begin{equation}
	T_{\rm accr} \approx  3300\ {\rm K}\ \left({\dot M\over 10^{-2}\ M_\oplus\ {\rm yr^{-1}}}\right)^{1/4}\left({M\over M_J}\right)^{1/4}\left({R\over 2R_J}\right)^{-3/4}.
\end{equation}
In the \cite{Marleau2017} simulations a constant equation of state was assumed so changes such as hydrogen dissociation are not considered. In addition, a 2D geometry, associated with disk accretion could cool the post-shock material as it spreads around the star, reducing the effective spherically-averaged temperature \citep{Hartmann1997, Geroux2016}. A recent discussion of the uncertainties in the thermodynamics of planetary accretion can be found in \cite{Mordasini2017}. Here, we follow \cite{Mordasini2013} and write 
\begin{equation}\label{eq:T0}
T_0^4 = \chi T_{\rm accr}^4 + T_{\rm eff}^4,
\end{equation}
where $T_{\rm eff} = (L/4\pi R^2\sigma)^{1/4}$ is the effective temperature, $L$ the internal luminosity of the planet, and $\chi$ is a parameter that accounts for the uncertainty in the shock temperature. We show results for two different choices, $\chi=1$ (as in BC17) and $\chi=0.1$, that bracket the optically-thick and thin results of \cite{Marleau2017}. We apply this temperature as an outer boundary condition during accretion, at a pressure corresponding to the ram pressure (the expected post-shock pressure) $P_0 = \dot M v_{ff}/4\pi R^2$, where the free fall velocity is given by $v_{ff}^2=2GM/R$. The temperature of the accreting material changes as it passes through the accreting envelope and becomes part of the planetary interior (\citealt{Berardo2017a}; see also \citealt{Stahler1988}). This is different from the approach of "hot accretion" in which energy is injected directly into the interior (e.g., \citealt{Kunitomo2017}). In fact, direct mixing into the convective interior is unrealistic for hot accretion because the accreted material forms a radiative zone around the convective core \citep{Geroux2016,Berardo2017a}.

The change of the accretion rate with time is important because it affects the shock conditions and therefore the entropy of the newly accreted matter. To explore the influence of the accretion rate history, we consider three cases. 
In the first one ({\em Case-A}) we implement the limiting gas accretion rate determined by the hydrodynamic simulations of \cite{Lissauer2009}. 
Their fitting formula written with $M_\oplus$ as the unit of mass and scaled to a maximum accretion rate $\dot M_{\rm max}$ is
\begin{eqnarray}\label{eq:mdot}
\log_{10}{\dot M\over \dot M_{\rm max}} &=& a_0 + a_1 \log_{10}{M\over M_\oplus} - a_2 \left[\log_{10}{M\over M_\oplus}\right]^2
\end{eqnarray}
where $a_0=-4.33$, $a_1=4.62$ and $a_2=1.23$.
With this form, $\dot M(t)$ rises to a peak value which we refer to as $\dot M_{\rm max}$ when $M\approx 75\,M_\oplus$ ($\log\, M/M_\oplus\approx 1.88$),
and then decreases again (Fig.~\ref{fig:formation_models}). 
In the second case  ({\em Case-B}), we use the same functional form as Case-A, but keep the accretion rate constant once it reaches its maximum value.
In the third case ({\em Case-C}), we 
follow Case-A up to $M=0.85\, M_J$, but then following \cite{Hubickyj2005}, linearly ramp down $\dot M$ by a factor of 100 over the remaining $0.15\, M_J$. In all cases, we scale the overall magnitude of the accretion rate, using $\dot M_{\rm max}$ as a parameter, but assume the functional form of $\dot M(M)$ remains the same. We find that our qualitative conclusions are robust when using other functional forms for $\dot M$ that rise to a peak and then fall (e.g., sin/cos). 

Fig.~\ref{fig:prof_1} shows the temperature and entropy profiles for different model assumptions when the protoplanet  reaches $1\,M_J$. The left panel shows models with the lowest luminosity and starting entropy ($L_{\rm in}=10^{-8}\ L_{\odot}$ and $S\approx 8\,k_B/m_u$), and accretion Case-B. These are the most favorable conditions for forming a radiative interior because the interior has the greatest entropy contrast with the accreted gas, and the increasing $\dot M$ leads to an increasing entropy of the accreted gas over time. The protoplanet's interior consists of layers of increasing entropy, suppressing convection. The models have a similar entropy profile, with a rapid outwards increase in the innermost layers from the initial value $S\approx 8\,k_B/m_u$ to $S\approx 9$--$10\,k_B/m_u$ depending on the assumed accretion rate. 
A higher shock temperature results in higher entropies in the outer regions. Only for the case with the lowest $\dot M$ and $\trat$ do we find a convective interior, but even this model has a radiative layer in the region above the core.

\begin{table*}
	\centering
\caption{Properties of the formation models at beginning of the detached phase, defined as the point where the gas accretion rate $\dot M$ becomes equal to the limiting accretion rate that can be supplied by the disk
(eq.~[\ref{eq:mdot}] with the maximum rate set to $\dot M_{\rm max}$ as given below). The second column gives the opacity relative to the Bell \& Lin (1994) opacity $\kappa_{BL}$. 
$S_{\rm inner}$ is the specific entropy in the inner part of the envelope. The last column gives the calculated time to reach crossover (i.e., $M_{\rm core}=M_{\rm env}$). }
\label{tab:formation_models}
\begin{tabular}{llllllll}
\hline
$\dot M_Z$ & $\kappa/$ & $M_{\rm core}$ & $R_{\rm core}$ & $M_{\rm env}$ & $\dot M$ & $S_{\rm inner}$ & $t_{\rm cross}$\\
$(M_\oplus/{\rm yr})$ & $\kappa_{BL}$ & $(M_\oplus)$ & $(R_\oplus)$ & $(M_\oplus)$ & $(M_\oplus/{\rm yr})$ & $(k_B/m_u)$ & $({\rm Myr})$\\
\hline
\multicolumn{7}{c}{$\dot{M}_{\rm max} = 10^{-3} M_\oplus/{\rm yr} $}\\
\hline
$10^{-7}$ & 0.01 & 4.87 & 1.5 & 24.6 & $6\times 10^{-4}$ &  8.56 & 27 \\ 
$10^{-7}$ & 0.1 & 8.2 & 1.75 & 30.4 & $7.8\times 10^{-4}$ & 9.08 & 30 \\
$10^{-6}$ & 0.01 & 8.2 & 1.75 & 21.0 & $6.0\times 10^{-4}$ & 8.84 & 3\\
$10^{-7}$ & 1 & 12.7 & 2.0 & 40.9 & $9.4\times 10^{-4}$ & 9.60 & 76\\
$10^{-6}$ & 0.1 & 12.7 & 2.0 & 30.3 & $8.7\times 10^{-4}$ & 9.42 & 7.6\\ 
$10^{-5}$ & 0.01 & 12.7 & 2.0 & 16.1 & $6.0\times 10^{-4}$ & 9.10 & 0.76\\
$10^{-6}$ & 1 & 19.0 & 2.3 & 41.3 & $1.0\times 10^{-3}$ & 9.95 & 14 \\
$10^{-5}$ & 0.1 & 19.0 & 2.3 & 25.0 & $8.7\times 10^{-4}$ & 9.70 & 1.4 \\
$10^{-5}$ & 1 & 26.0 & 2.57 & 33.4 & $1.0\times 10^{-3}$ & 10.2 & 2.2\\ 
\hline
\multicolumn{7}{c}{$\dot{M}_{\rm max} = 10^{-2} M_\oplus/{\rm yr} $}\\
\hline
$10^{-7}$ & 0.01 & 4.89 & 1.5 & 29.7 & $7.9\times 10^{-3}$ &  8.70 & 27\\ 
$10^{-7}$ & 0.1 & 8.2 & 1.75 & 35.0 & $9.0\times 10^{-3}$ & 9.18 & 30\\
$10^{-6}$ & 0.01 & 8.2 & 1.75 & 31.0 & $8.7\times10^{-3}$ & 9.10 & 3\\
$10^{-7}$ & 1 & 12.8 & 2.0 & 45.0 & $1.0\times 1 0^{-2}$ & 9.71 & 76\\
$10^{-6}$ & 0.1 & 12.8 & 2.0 & 44.0 & $1.0\times 10^{-2}$ & 9.66 & 7.6\\ 
$10^{-5}$ & 0.01 & 12.8 & 2.0 & 31.0 & $9.4\times 10^{-3}$ & 9.43 & 0.76\\
$10^{-6}$ & 1 & 19.1 & 2.3 & 52.0 & $1.0\times 10^{-2}$ & 10.1 & 14\\
$10^{-5}$ & 0.1 & 19.1 & 2.3 & 42.0 & $1.0\times 10^{-2}$ & 9.95 & 1.4\\
$10^{-5}$ & 1 & 27.0 & 2.6 & 53.0 & $1.0\times 10^{-2}$ & 10.4 & 2.2\\ 
		\hline
	\end{tabular}
\end{table*}

The size and location of the radiative region depends on the starting entropy and accretion rate history. As seen from the right panel of Fig.~\ref{fig:prof_1}, larger values of $L_{\rm in}$ lead to a higher entropy envelope initially, resulting in an inner convective region. A decreasing $\dot M$ with time leads to an outer convection zone because the accreted material arrives with a lower  entropy value that keeps decreasing. This is most effective for Case-C accretion which has the largest drop in $\dot M$, leading to more than half the mass becoming convective. For lower accretion rates with $\dot M_{\rm max} = 10^{-3}\ M_\oplus\ {\rm yr^{-1}}$, the inner and outer convection zones can merge leading to a fully-convective planet. For $\dot M_{\rm max} = 10^{-2}\ M_\oplus\ {\rm yr^{-1}}$ on the other hand, all cases have a radiative zone.
Overall we find that a proto-Jupiter which is not fully convective is a very common outcome.

\section{Constraints on proto-Jupiter's entropy from formation models}\label{method}

Above, we investigated the state of proto-Jupiter using various initial masses, entropies and accretion rates that are reasonable but are not guided by a formation model. Next, we use formation models to constrain the initial entropy, the initial core mass and the associated gas accretion rate. This narrows the parameter space of the possible combinations. The models are based on standard core accretion models where Jupiter forms in situ at 5 AU around a 1 M$_{\odot}$ star, and is embedded in a disk with the boundary conditions for the temperature and pressure of $T_{\text{out}}$ = 125 K  and $P_{\text{out}}$= 0.7 dyn/cm$^2$. Further details on the formation model can be found in \citet{Venturini16,Venhel17} and references therein.

\begin{figure}
\includegraphics[width=\columnwidth]{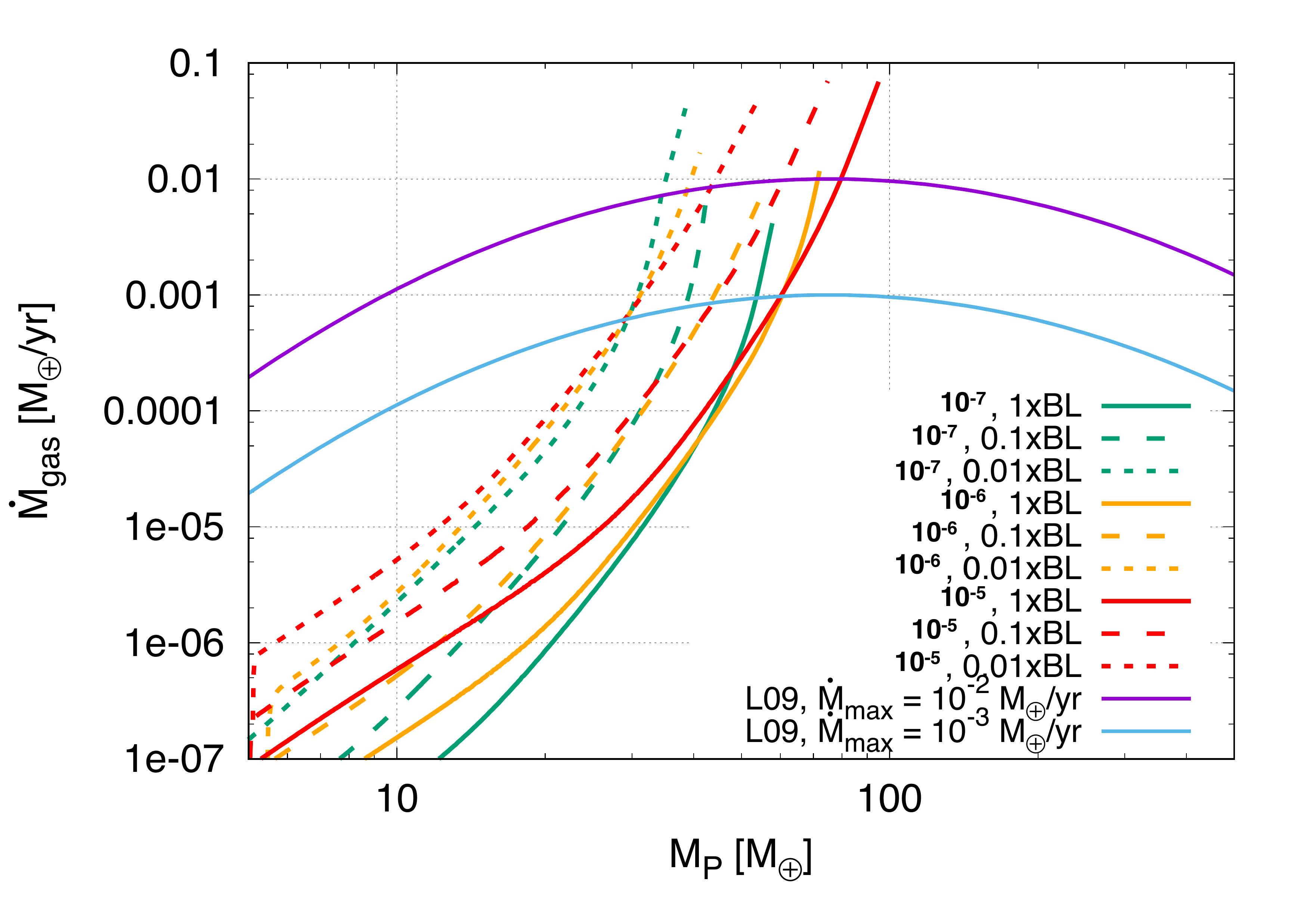}
\caption{Gas accretion rate vs.~planetary mass for the different formation models. The formation models assume the envelope is attached to the disk. When the gas accretion rate reaches the limiting gas accretion rate, determined by the ability of the disk to supply gas, the detached phase begins. To determine when this occurs, we use the limiting gas accretion rate from the calculations of \citet{Lissauer2009} (L09) (given in eq.~[\ref{eq:mdot}]), shown here for two different maximum rates $\dot M_{\rm max}$.} \label{fig:formation_models}
\end{figure}

\begin{figure*}
\includegraphics[width=1.015\columnwidth]{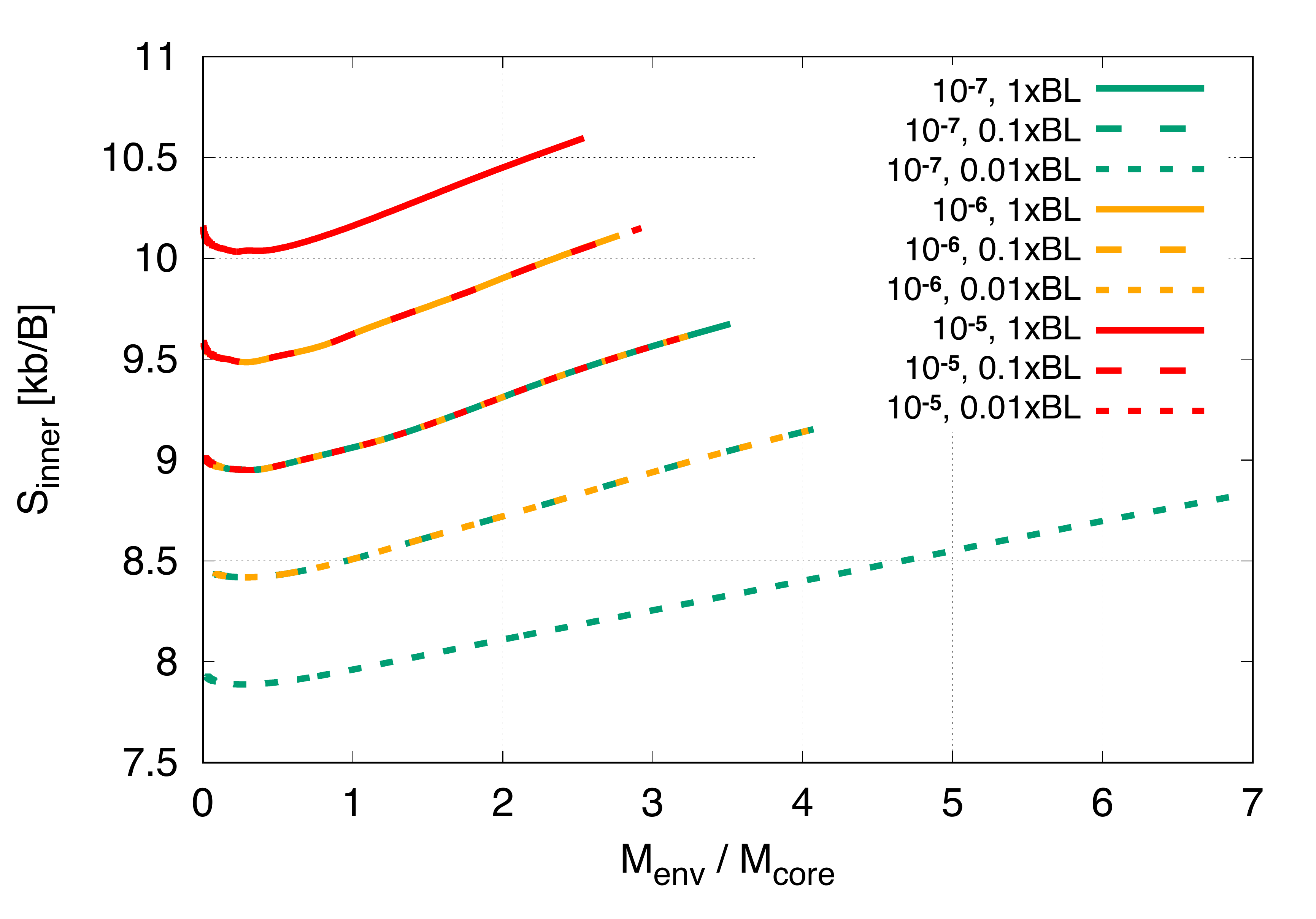}
\includegraphics[width=\columnwidth]{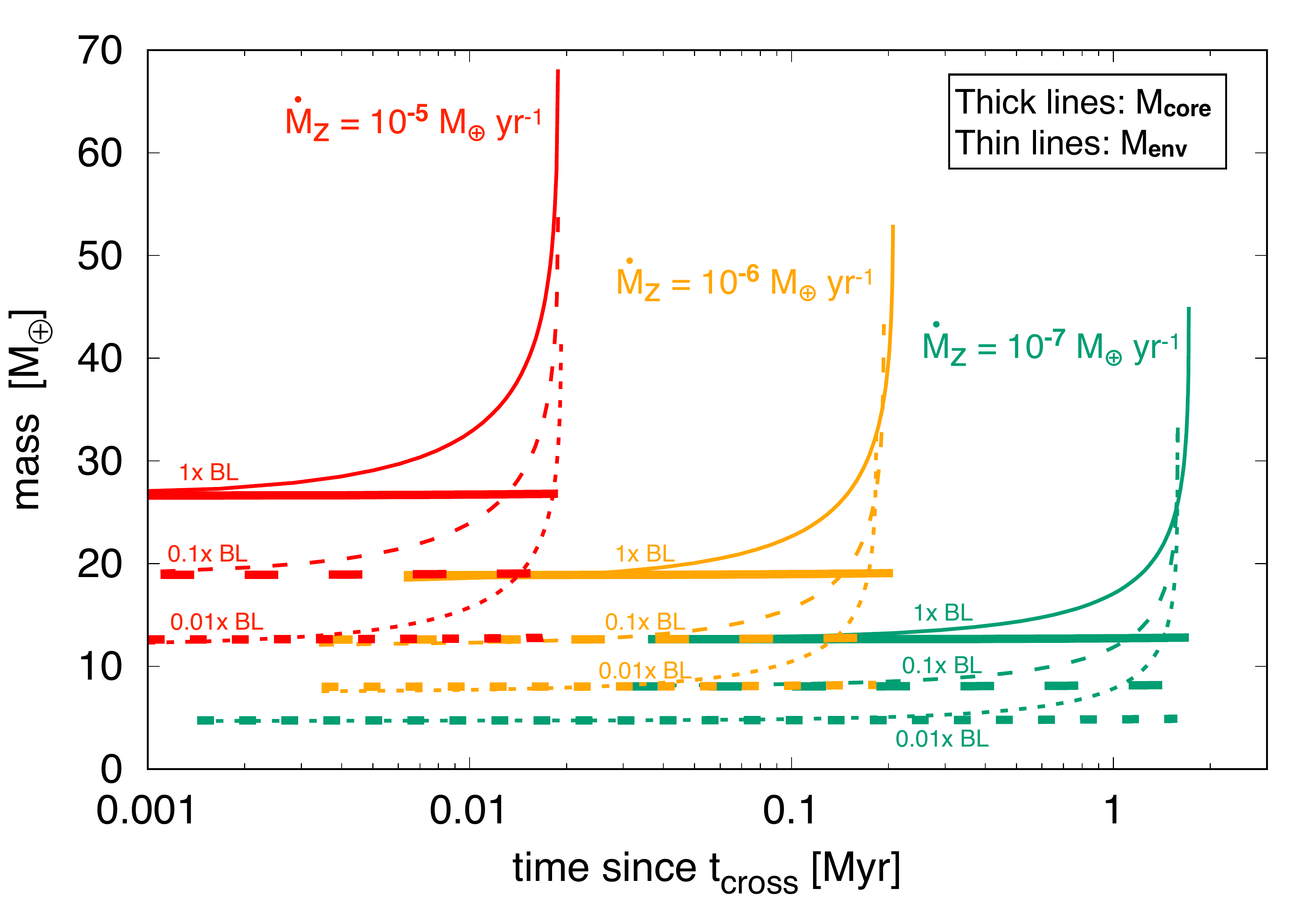}
\caption{{\bf Left:} Evolution of entropy at the innermost convective region of the envelope for the different formation models. The legend provides the value of $\dot M_Z/M_\oplus\,{\rm yr^{-1}}$ and the assumed opacity value. Varying accretion rate leads to different entropy profiles, but they are within the range of the ones we show here. {\bf Right:} The evolution of core mass (thick lines) and envelope mass (thin lines) as a function of time since crossover (see Table \ref{tab:formation_models} for values of $t_{\rm cross}$).} \label{fig:formation_models2}
\end{figure*}

\begin{figure*}
\includegraphics[width=2.1\columnwidth]{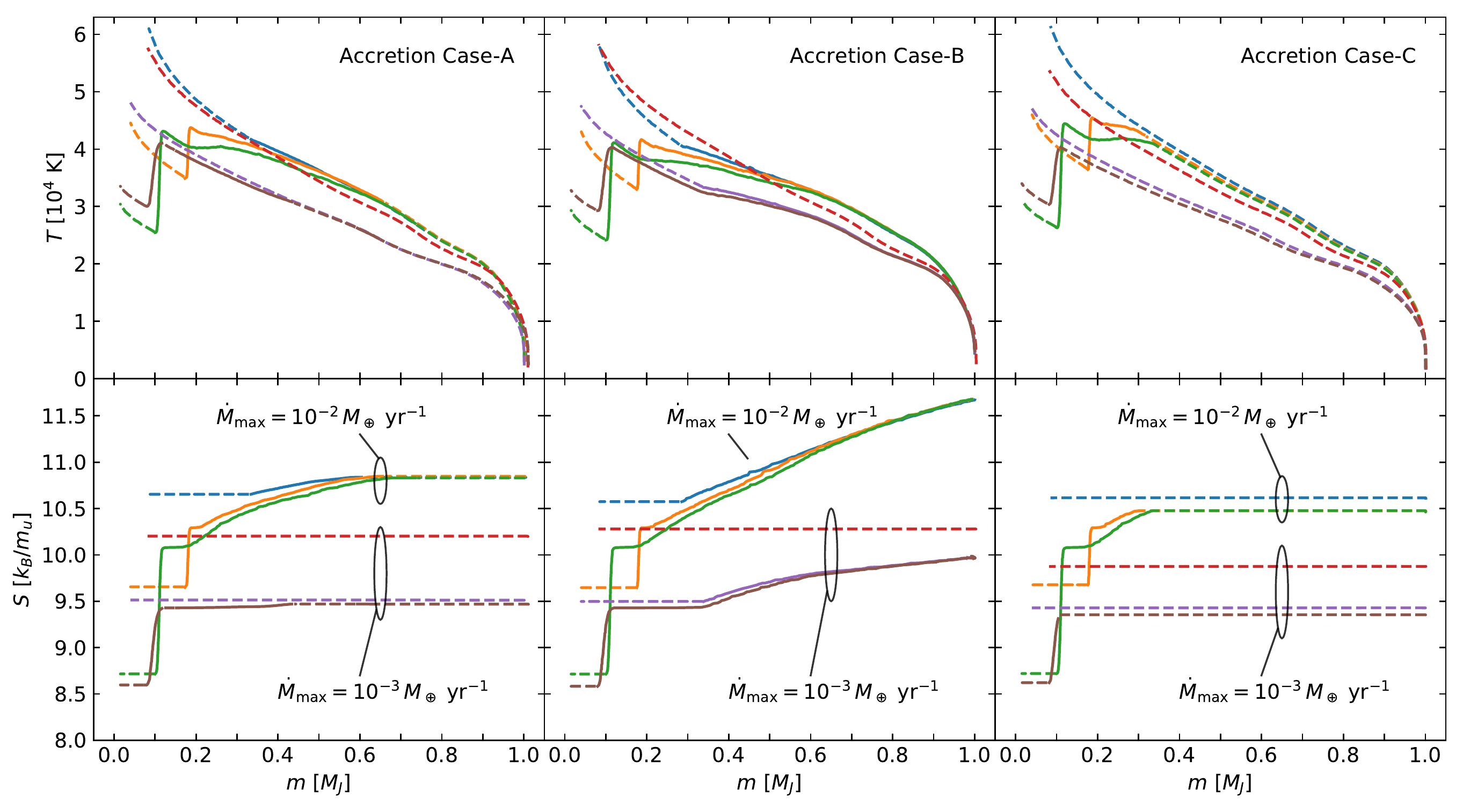}
\caption{Temperature and entropy profiles in proto-Jupiter for two different maximum accretion rates $\dot M_{\rm max}=10^{-3}$ and $10^{-2}\ M_\oplus\,{\rm yr^{-1}}$ and for three different starting models with $(\dot M_Z/M_\oplus\,{\rm yr^{-1}}, \kappa/\kappa_{BL})=$ $(10^{-7},0.01)$,  $(10^{-6},0.1)$ and $(10^{-5},1.0)$ (green, orange and blue curves for $\dot M_{\rm max}=10^{-2}\ M_\oplus\,{\rm yr^{-1}}$, and brown, purple, and red curves for $\dot M_{\rm max}=10^{-3}\ M_\oplus\,{\rm yr^{-1}}$, respectively). The three starting models have core masses $M_{\rm core}\approx 4.9$, $12.7$, and $26\,M_\oplus$, respectively. Convective regions are shown as dashed lines; solid lines are radiative zones. The models correspond to the three different accretion histories Case-A (left), Case-B (middle) and Case-C (right) assuming $\trat=1$.}\label{fig:prof_3}
\end{figure*}

We simulate Jupiter's formation for a range of opacities and solid accretion rates. A summary of the cases and the inferred parameters from the simulations are given in Table \ref{tab:formation_models}. We consider a large parameter space of possibilities. For the opacities we take as a baseline the classical opacities of \citet{Bell94} (hereafter BL94) with $\sim$ 1 g/cm$^2$ at $\sim$100 K. However, since it unclear which opacity values are most realistic for giant protoplanet atmospheres and how they may change with time due to solid accretion and grain coagulation and settling \citep{Movshovitz10,Mordasini14,Ormel14}, we also consider the BL94 opacities reduced by a factor of 10 and 100. For the solid accretion rate, also uncertain, we use three different values of $\dot M_Z=10^{-7}$, 10$^{-6}$, and 10$^{-5}$ \ME /yr. Clearly, the solid accretion rate is expected to change with time, but for simplicity we assume constant $\dot M_Z$ values. The solid accretion rate near the onset of the detached phase is most important since it determines the temperature and entropy of the envelope at the time when the planet becomes detached and starts to accrete through a shock. 
The time to reach crossover mass $t_{\rm cross}$ (i.e.~the time when $M_{\rm core}=M_{\rm env}$) is given in Table~\ref{tab:formation_models}. Because it is linked to the assumed constant $\dot M_Z$, it is not necessarily realistic.    
For example, for the models with low solid accretion rate of $\dot M_z=10^{-7}$ \ME /yr, the crossover time is longer than the expected disk lifetime. However, these models could represent scenarios in which first the core forms rapidly by pebble accretion, followed by a slower accretion of planetesimals.
 
Fig.~\ref{fig:formation_models} shows the calculated gas accretion rate vs.~planetary mass for the different formation models.  Once runaway gas accretion begins, $\dot M$ grows rapidly and eventually reaches the maximum rate that can be supplied by the disk (the detached phase). To identify this point in our simulations, we compared the accretion rate from the formation model with the limiting disk accretion rate from \cite{Lissauer2009} normalized to a given maximum rate $\dot M_{\rm max}$ (both are shown in Fig.~\ref{fig:formation_models}). Table \ref{tab:formation_models} lists the properties of the model at the beginning of the detached phase for two different values of $\dot M_{\rm max}$. 

The left panel of Fig.~\ref{fig:formation_models2} shows the entropy of the innermost convective region of the envelope ($S_{\text{inner}}$). At early stages, the planetary entropy is low and constantly increases as more gas is accreted and the planet gains mass. The entropy depends on the product of opacity and solid accretion rate. A decrease in opacity by a factor of ten, for example, is compensated by an increase in $\dot M_z$ by a factor of ten, so that several of the entropy curves overlap. This degeneracy arises because the cooling rate of the envelope for a given entropy is set by the radiative gradient at the radiative-convective boundary $\propto L \kappa \propto \dot M_z \kappa$.
In reality, both parameters are expected to change with time, and opacity can change by a different factor than $\dot M_z$. Therefore the entropy's time evolution will be more complex than shown in Figure \ref{fig:formation_models2}.
The time-evolution of the core and envelope masses after crossover is shown in the right panel of the figure. As expected, higher solid accretion rate leads to more massive cores, and more rapid planetary growth. Higher opacities result in slower growth and therefore a higher core mass for a given $\dot M_z$.    

Clearly, the growth history determines the planetary entropy, core mass, and gas accretion rate at the onset of the detached phase. Typically, protoplanets with small cores have lower entropies and lower gas accretion rates. Note that low solid accretion rates (and therefore, $L_{\rm in}$ and entropy) at the detached phase are expected for accretion of both pebbles and planetesimals. If the protoplanet grows primarily by pebbles, they are likely to dissolve in the upper envelope; while if mostly planetesimals are accreted, only large ones reach the core, and their accretion is very inefficient \citep{IP07,fortier13}. Therefore, a low luminosity above the core, and the formation of a radiative region is a likely outcome. 

We used the values of core and envelope masses and entropy for six cases from Table \ref{tab:formation_models} as initial conditions for MESA models of the detached phase. For consistency, our formation models also assume a H-He envelope with no heavy elements. Therefore the entropy from the formation models can be used as an input for modeling the last stages of the planetary formation. The resulting temperature and entropy profiles of proto-Jupiter are shown in Fig.~\ref{fig:prof_3}, for each of the three different accretion cases and $\trat=1$. 

We find that a range of different outcomes is possible, with some cases being fully-convective and others with significant radiative regions near the core. The models with the largest opacity $\kappa/\kappa_{BL}=1$ and solid accretion rate $\dot M_Z=10^{-5}\ M_\oplus\,{\rm yr^{-1}}$ end up being fully-convective, because the starting entropy is high $S_{\rm inner}\sim 10\, k_B/m_u$, comparable to the entropy of the accreted material. Lower accretion rates and opacities lead to a more complicated internal entropy profile, and the existence of radiative regions. At the higher gas accretion rate $\dot M_{\rm max}=10^{-2}\ M_\oplus\,{\rm yr^{-1}}$, an extended radiative region separates inner and outer convection zones; for one of the $\dot M_{\rm max}=10^{-3}\ M_\oplus\,{\rm yr^{-1}}$ models, the transition is sharper and the planet has a two-layer convection zone structure. We see that a variety of entropy profiles can be created depending on the opacity of the envelope and accretion rate of solids; the phase of planet formation before runaway accretion leaves its mark on the final internal profile.

\begin{figure}
\includegraphics[width=1.02\columnwidth]{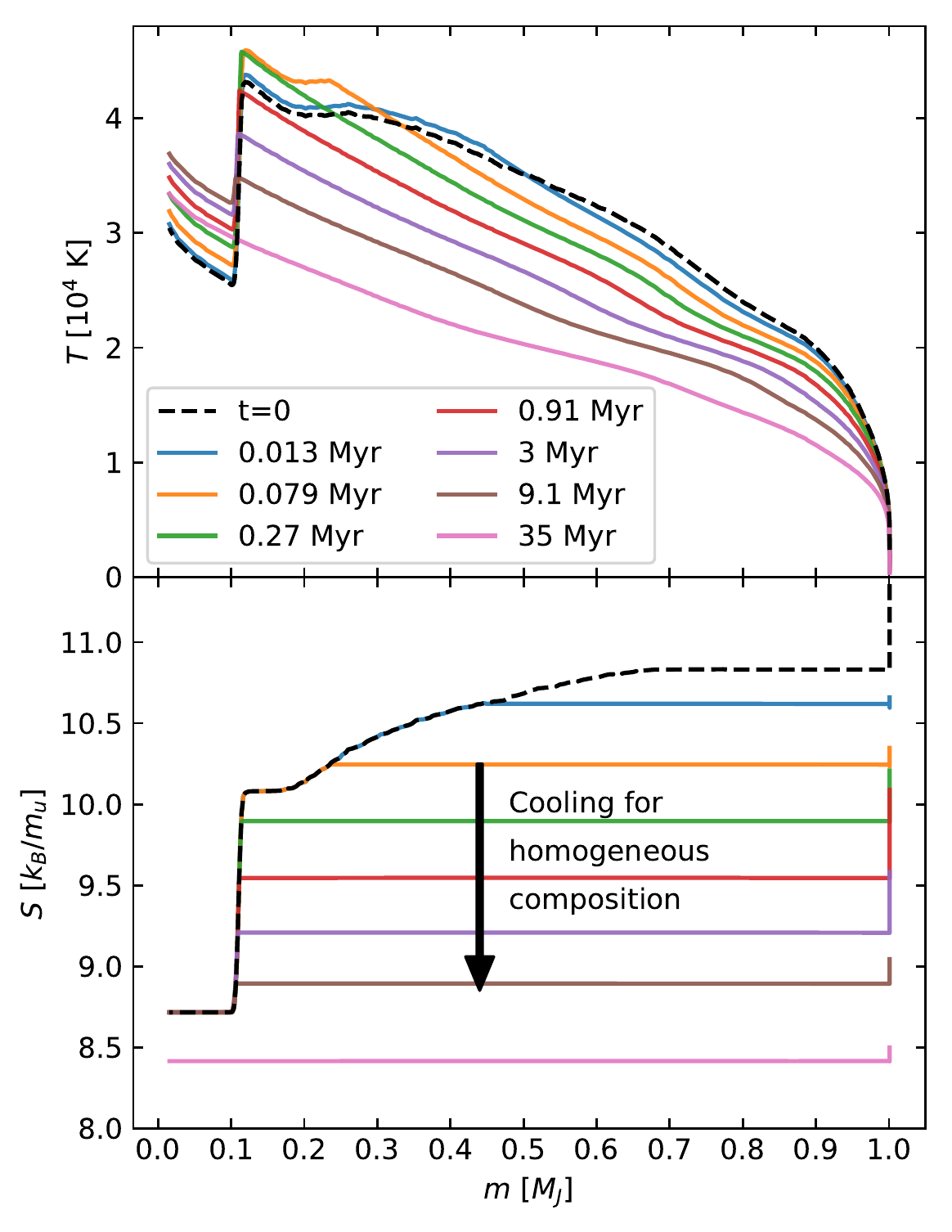}
\caption{The evolution of the temperature and entropy profile during post-accretion cooling. The initial profile (dashed curve) is a model from Fig.~\ref{fig:prof_3} with a large entropy contrast ($\dot M_Z/M_\oplus=10^{-7}\,{\rm yr^{-1}}$, $\kappa/\kappa_{BL}=0.01$, and $\dot M_{\rm max}=10^{-2}\ M_\oplus\,{\rm yr^{-1}}$, accretion Case-A). As the planet cools, the outer convection zone penetrates inwards. The entropy barrier between innermost layers and the outer envelope is erased after a few Myr.}\label{fig:prof_cool}
\end{figure}

\begin{figure}
\includegraphics[width=0.99\columnwidth]{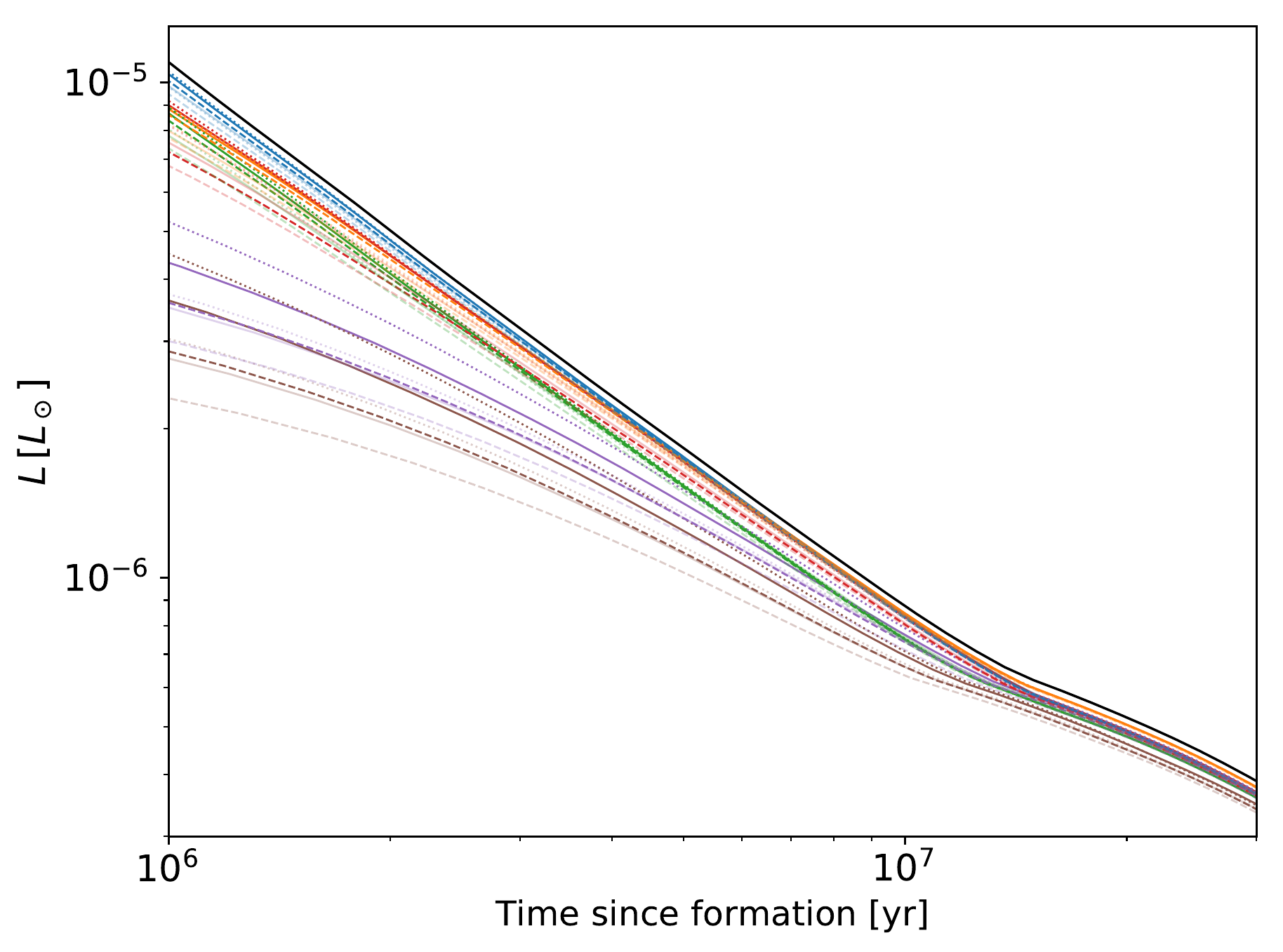}
\caption{Luminosity of the models shown in Figure \ref{fig:prof_cool} between 1 and $30\ {\rm Myr}$ after formation assuming a grey atmosphere and no core luminosity. The colors match those in Figure \ref{fig:prof_cool}. These models have $\trat=1$, and we indicate accretion case-A, case-B, and case-C with solid, dotted and dashed lines, respectively. To illustrate the effect of the shock temperature, we include a fainter set of curves showing the results for the same set of parameters but with $\trat=0.1$. For comparison, the black solid curve shows a ``hot start'' fully-convective model with starting entropy $S_i\approx 11\ k_B/m_p$. 
}\label{fig:lum}
\end{figure}

\section{Discussion and Conclusions}

By coupling formation and accretion models, we have calculated the primoridal entropy profile of Jupiter. The result depends on both the early stages of planetary growth and the details of the later phases of runaway gas accretion. Our main findings are:
\begin{itemize}
\item Lower opacity and lower solid accretion rate prior to detachment lead to a low mass core and a low entropy in the gas envelope. 
\item The contrast with the entropy of the accreted gas can then lead to an extended radiative zone in the inner regions. 
\item Higher accretion rates and shock temperatures increase this contrast and the resulting entropy gradient.
\item The rate at which the accretion rate drops as the planet reaches its final mass determines how far inwards the outer convection zone is able to penetrate. 
\item During the detached phase, if the gas accretion rate and the shock temperature are high the protoplanet is likely to consist of a radiative region. 
\item A fully-convective interior forms when the contrast between the interior entropy and accreted entropy is small. This typically occurs with low gas accretion rate or shock temperature, high solid accretion rate or opacity, and when the gas accretion rate turns off slowly.
\end{itemize}

It will be important to include the internal entropy profiles that we find here in future studies to determine the evolution and current-state of Jupiter.  The thermal stratification could influence the distribution of heavy elements, for example by delaying mixing of heavy elements from the innermost regions near the core into the outer convective envelope. In Figure \ref{fig:prof_cool}, we show how the entropy profile evolves during cooling for one of our models. With the homogeneous composition we assume here, cooling leads to a fully-convective interior in $\sim 10^7\ {\rm yr}$. 
This could have an important effect on core erosion. 
The estimates of \cite{Guillot2004} suggest that a substantial fraction of the erosion occurs at young ages, when the cooling luminosity is larger and convection is more vigorous. 
In the model shown in Figure \ref{fig:prof_cool}, core erosion is delayed by $\sim 10^7\ {\rm yr}$, the time it takes to overcome the entropy barrier at $m\approx 0.1\ {\rm M_J}$. 
More detailed investigations of the influence of the entropy profile on core erosion and mixing in giant protoplanets are required. In addition, it is desirable to include solid accretion during runway and investigate how it affects the final composition, internal structure and long-term evolution. 

Composition gradients could significantly change the cooling shown in Figure \ref{fig:prof_cool}, by delaying or even preventing the planet from becoming fully-convective. With heavy elements added, the radiative regions laid down by accretion could persist so that Jupiter's interior may not be fully-convective today. \cite{vazan18} recently explored initial models for Jupiter with composition gradients that remain non-adiabatic today and still satisfy the observational constraints on Jupiter's interior. Interestingly, the shape of the entropy profile (i.e.~the contrast in entropy between the inner and outer regions) we find here is very similar to the one derived by \citet{vazan18}, although the overall value of entropy is lower because of the high metal content of the models. A Jupiter with a diluted core up to $\sim$ 50\% of Jupiter's mass is a suggested model for Jupiter structure which is consistent with the Juno data \citep{wahl17}. Further models of formation and evolution including heavy elements are needed. Seismology would be another way to probe the presence of stable regions in Jupiter's interior (e.g.~\citealt{Gaulme2014}).

Our results have implications for the characterization of young Jupiter-mass exoplanets detected by direct imaging. The internal entropy of a newly-formed gas giant determines the planet's luminosity at young ages \citep{Marley2007,Spiegel2012}. In the core accretion framework, we find a large range of primordial entropies spanning $\approx 8$--$11\,k_B/m_u$, and corresponding  luminosities. Figure \ref{fig:lum} shows the luminosity at early times between $1$ and $30\ {\rm Myr}$ after formation, when the differences between models are most pronounced. For comparison, we also show a fully-convective ``hot start'' model, commonly used to determine the masses of directly-imaged planets. 
Figure \ref{fig:lum} shows that the luminosity of a Jupiter-mass planet can vary by factors of 2 to 3 depending on the formation history (see also \citealt{Mordasini2017}).
We note that the hot start model is lower by $\approx 30$--$50$\% compared to the model of \cite{Burrows1997} which includes detailed non-grey atmospheres. Therefore inferring the planetary mass from luminosity in fact depends on the planetary entropy and atmospheric properties.
Finally, it has been suggested that luminosity could distinguish planets formed by core accretion from those formed by disk instability (e.g.~see discussion in \citealt{Mordasini2012}). Figure \ref{fig:lum} shows that high-luminosity giant planets could form by core accretion so that hot vs.~cold (high-entropy vs.~low-entropy) start does not distinguish among these formation scenarios.

\section*{Acknowledgements}

A.~C.~is supported by an NSERC Discovery grant, is a member of the Centre de Recherche en Astrophysique du Qu\'ebec (CRAQ), and thanks the School of Mathematics, Statistics and Physics at Newcastle University for hospitality. R.~H.~acknowledges support from SNSF grant 200021\_169054. Part of this work  was conducted  within the framework of the National Centre for Competence in Research PlanetS, supported by the Swiss National Foundation.

\bsp	
\label{lastpage}
\end{document}